\documentclass[conference]{IEEEtran}
\IEEEoverridecommandlockouts

\usepackage{cite}
\usepackage{amsmath,amssymb,amsfonts,bm}
\usepackage{algorithmic}
\usepackage{graphicx}
\usepackage{textcomp}
\usepackage{xcolor}
\def\BibTeX{{\rm B\kern-.05em{\sc i\kern-.025em b}\kern-.08em
    T\kern-.1667em\lower.7ex\hbox{E}\kern-.125emX}}
\usepackage{graphicx}
\usepackage{booktabs}
\usepackage{multirow}
\usepackage{adjustbox}
\usepackage{amssymb}
\usepackage{pifont}
\newcommand{\cmark}{\ding{51}}
\newcommand{\xmark}{\ding{55}}
\usepackage{xcolor}
\usepackage{hyperref}
\usepackage{url}

\begin{document}

\title{LAVCap: LLM-based Audio-Visual Captioning using Optimal Transport}
\author{\begin{tabular}{c}
\textit{Kyeongha Rho$^{1*}$, Hyeongkeun Lee$^{1*}$, Valentio Iverson$^2$, Joon Son Chung$^1$}\\
$^1$KAIST, South Korea,
$^2$University of Waterloo, Canada \\
\{khrho325, lhk528, joonson\}@kaist.ac.kr, viverson@uwaterloo.ca
\end{tabular}
\thanks{$^*$These authors contributed equally to this work.}
}

\maketitle

\begin{abstract}
Automated audio captioning is a task that generates textual descriptions for audio content, and recent studies have explored using visual information to enhance captioning quality. However, current methods often fail to effectively fuse audio and visual data, missing important semantic cues from each modality.
To address this, we introduce LAVCap, a large language model (LLM)-based audio-visual captioning framework that effectively integrates visual information with audio to improve audio captioning performance.
LAVCap employs an optimal transport-based alignment loss to bridge the modality gap between audio and visual features, enabling more effective semantic extraction. Additionally, we propose an optimal transport attention module that enhances audio-visual fusion using an optimal transport assignment map.
Combined with the optimal training strategy, experimental results demonstrate that each component of our framework is effective.
LAVCap outperforms existing state-of-the-art methods on the AudioCaps dataset, without relying on large datasets or post-processing. Code is available at \url{https://github.com/NAVER-INTEL-Co-Lab/gaudi-lavcap}.
\end{abstract}

\begin{figure*}[t!]
    \centering
    \includegraphics[width=0.95\textwidth]{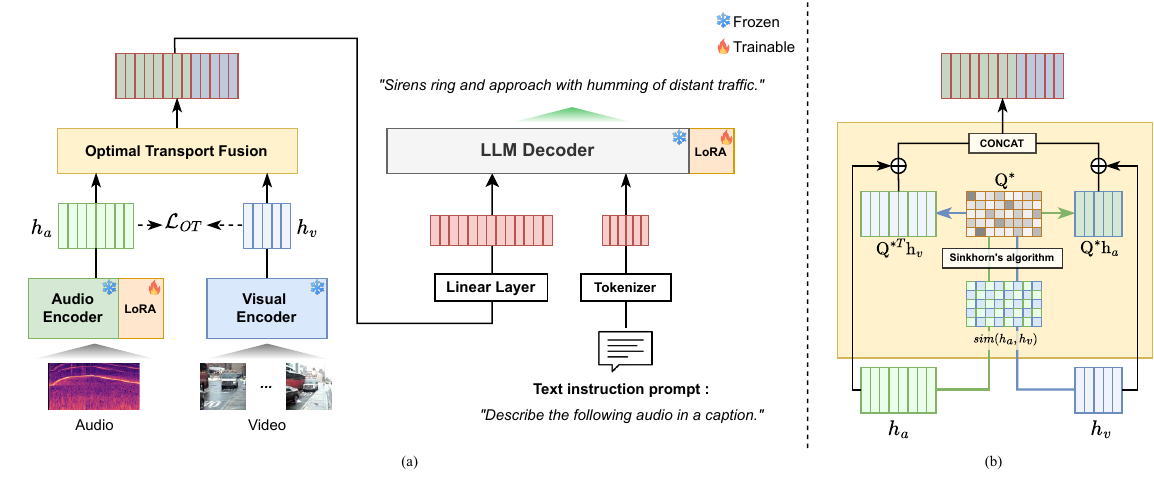}
    \vspace{-1.3em}
    \caption{(a) Overview of the proposed LAVCap Framework. (b) Detail of the Optimal Transport Fusion module.}
    \label{fig:main_framework}
    \vspace{-1.4em}
\end{figure*}

\section{Introduction}

Automated audio captioning (AAC) is a task that aims to generate textual descriptions of a given audio input. 
In contrast to audio tagging and sound event classification tasks that focus on understanding low-level sound characteristics, AAC also requires a comprehension of high-level semantics contained in the audio.
It has recently garnered significant attention from the deep learning community due to its wide range of applications, such as providing audio descriptions for broadcasts and movies, text query-based audio retrieval, and developing more human-like conversational AI systems~\cite{huang2023personalized}.

Most recent works~\cite{mei2021act, gontier2021barttags,kim2023prefix,kim2024enclap,mei2024wavcaps,labb2024conette,liu2024enhancing,haji2024taming} for AAC adopt the encoder-decoder framework and have explored different architectures.
They utilize various audio encoders~\cite{kong2020panns,chen2022hts,chen2023beats,elizalde2023clap,défossez2023high} to extract semantically rich acoustic features.
Pre-trained language models are used as text decoders because of their sequence modeling and text generation capabilities.
In particular, some recent works~\cite{kim2023prefix, liu2024enhancing} leverage LLMs for their strong ability to understand contexts and generate text sequences.

Recently, several studies~\cite{liu2023visually, kim2024avcap} have incorporated visual modality to improve audio captioning.
Visual information helps distinguish sounds in complex scenes.
For example, when a man speaks while drilling, visual cues clarify the presence of both elements, allowing for generating more accurate captions.

To this end, we introduce \textbf{LAVCap}, an \textbf{L}LM-based \textbf{A}udio-\textbf{V}isual \textbf{Cap}tioning framework.
This work focuses on exploring various audio-visual fusion methods and identifying the optimal training strategy for the LLM-based audio-visual captioning framework.
First, we find out that naively combining visual features with audio features does not enable the LLM to take full advantage of the visual information.
We attribute this shortfall to the significant modality gap between audio and visual feature spaces.

To bridge the modality gap, we consider the alignment of audio and visual tokens as an optimal transport (OT) problem. While previous work \cite{lin2024norton} applies OT algorithms to align video and text, this study pioneers their application to audio-visual alignment. Specifically, we introduce an optimal transport-based alignment loss (OT loss) tailored for this purpose.

OT loss encourages the encoders to extract features that are rich in semantics while ensuring that they are well-aligned.
In addition, our experiment reveals that the existing cross-attention mechanism~\cite{liu2023visually} struggles to integrate cross-modal features effectively.
To tackle this, we propose an optimal transport attention module (\textit{OT-Att}) that leverages the OT assignment map as attention weights for audio-visual fusion. 
This approach provides more effective fusion compared to other audio-visual fusion methods.

Experimental results demonstrate that LAVCap outperforms previous state-of-the-art methods on the AudioCaps benchmark, without pre-training the model on large datasets or applying post-processing to generated captions.
It is noteworthy that our approach achieves such high performance although the ground-truth captions in AudioCaps are audio-centric.
Furthermore, our user study shows that LAVCap obtains mean opinion scores (MOS) even higher than the ground-truth captions.
This underscores the importance of utilizing the visual modality to distinguish the various sounds and understand the scene.

The main contributions of this work are summarized as follows:
\begin{itemize}
    \item We present \textbf{LAVCap}, an LLM-based audio-visual captioning framework that leverages visual information to complement audio modality in audio captioning tasks.
    \item We employ optimal transport for effectively bridging the modality gap and fusing audio-visual features.
    \item Our approach outperforms existing state-of-the-art methods on AudioCaps without the need for pre-training the model on large datasets or applying post-processing.
\end{itemize}

\section{method}
The overall architecture of LAVCap is illustrated in Fig.~\ref{fig:main_framework}.
Each component of our framework and the training objectives are explained in detail in the following sections.

\subsection{Audio-Visual Encoding}
Given an audio-visual input pair ($x_a$, $x_v$), the audio encoder $E_a$ and the visual encoder $E_v$ encode them into modality-specific features with $C$ dimensions as follows:
\begin{equation}
    h_a = E_a(x_a),\;  h_v = E_v(x_v)
\end{equation}
where $h_a \in \mathbb{R}^{N_a \times C}$, $h_v \in \mathbb{R}^{N_v \times C}$ represent the audio and visual features, respectively, with $N_a$ and $N_v$ denoting the number of tokens for audio-visual features.

\subsection{Audio-Visual Alignment Based on Optimal Transport}
To effectively bridge the modality gap, we apply OT loss to align the audio-visual features.
First of all, we compute the similarity matrix $\mathbf{S} \in \mathbb{R}^{N_a \times N_v}$ of audio and visual features.
Similar to the prior work~\cite{lin2024norton}, we optimize the audio-visual assignment map $\mathbf{Q} \in \mathbb{R}^{N_a \times N_v}$ to maximize the global similarity between the cross-modal features:
\begin{equation}
\begin{split}
    &  \max_{\mathbf{Q} \in \mathcal{Q}} \quad \text{tr}(\mathbf{Q}^\top \mathbf{S}) + \epsilon H(\mathbf{Q}) \\
    \text{s.t.} \,\, \mathcal{Q} = \{ \mathbf{Q} & \mid \mathbf{Q}\mathbf{1}_{N_v}=\frac{1}{N_a}\mathbf{1}_{N_a}, \, \mathbf{Q}^\top \mathbf{1}_{N_a}=\frac{1}{N_v}\mathbf{1}_{N_v} \},
\end{split}
\end{equation}
where $\mathbf{1}_{N_a}$, $\mathbf{1}_{N_v}$  denotes the vector of ones in dimension $N_a$ and $N_v$.
$H(\mathbf{Q})$ is an entropy regularization term with $\epsilon$ controlling the smoothness.
Then the optimal solution $\mathbf{Q}^*$ is obtained by iterative Sinkhorn-Knopp algorithm~\cite{cuturi2013sinkhorn}:
\begin{equation}
    \mathbf{Q}^* = \text{Diag}(\bm{\mu})\text{exp}(\mathbf{S}/\epsilon)\text{Diag}(\bm{\nu}) \;\;(\bm{\mu}\in\mathbb{R}^{N_a},\;\bm{\nu}\in\mathbb{R}^{N_v})
\end{equation}
where $\bm{\mu}$ and $\bm{\nu}$ are the non-negative scaling vectors.

Based on the optimal solution $\mathbf{Q}^*$ and the similarity matrix $\mathbf{S}$, OT loss is computed as follows:

\begin{equation}
    \mathcal{L}_\text{OT} = \sum_{m\in\{a,v\}}\Biggl[-\sum^{N_m}_{i=1}\log{\frac{ \exp{(\langle\mathbf{q}^m_i,\mathbf{s}^m_i\rangle/\tau)}}{\sum^{N_m}_{j=1} \exp{(\langle\mathbf{q}^m_i,\mathbf{s}^m_j\rangle/\tau )}}}\Biggr],
\end{equation}
\\
\noindent where $\{\mathbf{q}^m_{i}\}^{N_m}_{i=1}$, $\{\mathbf{s}^m_{i}\}^{N_m}_{i=1}$ denote the row vectors of $\mathbf{Q}$ and $\mathbf{S}$ when $m=a$, and the column vectors when $m=v$.
OT loss encourages maximizing the similarity between audio-visual token sequences within a sample, rather than between averaged features in a mini-batch.
This provides more fine-grained supervision and thus a more effective way to align the cross-modal features. 

\begin{table*}[t!]
\setlength{\tabcolsep}{10pt}
    \caption{AAC results on AudioCaps test split for various models. Bold indicates the best among audio-visual methods, while underlined represents the overall top performers. Note that ground-truth captions are based solely on audio. $^{\dagger}$results reproduced on our environment. AS: AudioSet, AC: AudioCaps, WC: WavCaps, CL: Clotho, MA: Multi-Annotator Captioned Soundscapes.}
    \vspace{-0.5em}
    \centering
    \begin{adjustbox}{max width=\textwidth}
    \begin{tabular}{lcccccccc}
        \toprule
        Model & Pre-training Dataset & $\text{BLEU}_1$ $\uparrow$ & $\text{BLEU}_4$ $\uparrow$ & $\text{ROUGE}_L$ $\uparrow$ & METEOR $\uparrow$& CIDEr $\uparrow$& SPICE $\uparrow$ & SPIDEr $\uparrow$ \\
        \midrule
        \small\textbf{\textit{Audio-Based}} \\
        ACT~\cite{mei2021act}& AS & 64.7 & 25.2 & 46.8 & 22.2 & 67.9 & 16.0 & 42.0 \\
        $\text{BART-}_\text{tags}$~\cite{gontier2021barttags}& AS & 69.9 & 26.6 & 49.3 & 24.1 & 75.3 & 17.6 & 46.5 \\
        CNN-GPT2~\cite{kim2023prefix} & - & 71.3 & 30.9 & 50.3 & 24.0 & 73.3 & 17.7 & 45.5 \\
        EnCLAP\small{-large}~\cite{kim2024enclap}  & - & - & - & - & 25.5 & 80.3 & 18.8 & 49.5 \\
        HTSAT-BART~\cite{mei2024wavcaps}  & AC+CL+WC & 70.7 & 28.3 & 50.7 & 25.0 & 78.7 & 18.2 & 48.5 \\
        CNext-trans~\cite{labb2024conette}   & AC+CL+MA+WC & - & - & - & 25.2 & 80.6 & 18.4 & 49.5 \\
        LOAE$^{\dagger}$~\cite{liu2024enhancing} & - & 69.8 & 25.8 & 49.2 & 24.8 & 75.6 & 18.2 &  46.9   \\
        LOAE~\cite{liu2024enhancing} & AC+CL+WC & - & - & - & \underline{26.7} & 81.6 & \underline{19.3} & 50.5    \\
        AutoCap (audio+text)~\cite{haji2024taming} & AC & 72.1 & 28.6 & 51.5 & 25.6 & 80.0 & 18.8 & 49.4    \\
        AutoCap (audio+text)~\cite{haji2024taming} & AC+CL+WC & \underline{72.3} & \underline{29.7} & \underline{51.8} & 25.3 & 83.2 & 18.2 &  50.7   \\
        \midrule
        \small\textbf{\textit{Audio-Visual}} \\
        V-ACT\cite{liu2023visually} & - & 69.8 & 28.1 & 49.4 & 23.7 & 71.1 & 17.2 & 44.2\\
        AVCap (freeze)\cite{kim2024avcap} & - & 70.8 & 29.5 & 49.8 & 22.8 & 74.4 & 16.2 & 45.5\\
        AVCap (finetuning)\cite{kim2024avcap} & - & 68.1 & 28.7 & 49.1 & 24.3 & 75.8 & 17.8 & 46.8\\
        LAVCaps (ours)  & - & \underline{\textbf{72.3}} & \underline{\textbf{29.7}} & \textbf{51.0} & \textbf{26.2} & \underline{\textbf{84.9}} & \textbf{18.5} & \underline{\textbf{51.7}}  \\
        \bottomrule
    \label{tab:main_tab}
  \end{tabular}
  \end{adjustbox}
  \vspace{-2em}
\end{table*}

\begin{table}[t!]
\setlength{\tabcolsep}{10pt}
    \vspace{-1em}
    \caption{Ablation on the use of visual modality and $\mathcal{L}_\text{OT}$.}
    \vspace{-0.5em}
    \centering
    \begin{adjustbox}{max width=\columnwidth}
    \begin{tabular}{ccccccc}
        \toprule
        Audio & Visual & $\mathcal{L}_\text{OT}$ & METEOR $\uparrow$ & CIDEr $\uparrow$ & SPICE $\uparrow$ & SPIDEr $\uparrow$ \\
        \midrule
        \cmark & \xmark & \xmark  & 24.6 & 77.9 & 17.4 & 47.6   \\
        \cmark & \cmark & \xmark  & 25.4 & 78.1 & 18.8 & 48.5  \\
        \cmark & \cmark & \cmark  & \textbf{26.3}  & \textbf{83.1} & \textbf{18.9} & \textbf{51.0}  \\
        \bottomrule
    \label{tab:abl_split}
  \end{tabular}
  \end{adjustbox}
  \vspace{-1em}
\end{table}

\subsection{Audio-Visual Fusion and Projection}
In addition to using the optimal transport assignment map for loss computation, we also employ it for the fusion of audio-visual features.
We refer to this fusion module as \textbf{OT-Att} module in Eq.~\ref{eq:ot-att}. OT-Att module operates similarly to the conventional cross-attention but utilizes the optimal transport assignment map as the attention weight.
The visually-attended audio feature and audio-attended visual feature are computed as follows:
\begin{equation}
\label{eq:ot-att}
\begin{split}
    \hat{h}_a & = \operatorname{\textbf{OT-Att}}(h_a, \mathbf{Q}^*, h_v) = h_a + \mathbf{Q}^* h_v \\
    \hat{h}_v & = \operatorname{\textbf{OT-Att}}(h_v, \mathbf{Q}^{*\top}, h_a) = h_v + \mathbf{Q}^{*\top} h_a
\end{split}
\end{equation}
Then these two features are concatenated token-wise and then projected to the LLM latent space through the linear projector:
\begin{equation}
    h_{av} = W_{av \rightarrow t}\operatorname{Concat}(\hat{h}_a, \hat{h}_v)
\end{equation}

\subsection{Text Decoding}
Since we leverage an LLM as the text decoder, the text instruction prompt $x_t$ needs to be transformed into text embeddings through the LLM tokenizer $E_{t}$.
Then the fused audio-visual features and text embeddings are concatenated and fed into the LLM decoder $D_t$ to produce the output $z_t$:
\begin{equation}
    z_{t} = D_t(\operatorname{Concat}(h_{av}, E_{t}(x_t)), M_{att})
\end{equation}
where $M_{att}$ denotes attention masks for reflecting the auto-regressive property of LLM.

\subsection{Training Objectives}
As well as OT loss, the conventional autoregressive cross-entropy loss is also used for training:
\begin{equation} 
    \mathcal{L}_{\text{CE}} = -\frac{1}{T} \sum_{i = 1}^T \log{p(y_i \mid y_{1:i-1}, h_{av}, h_t)}
\end{equation}
where $y_i$ is a $i$-th text token.
The final training objective is the weighted sum of two losses:
\begin{equation}
    \mathcal{L} = \lambda_{CE}\mathcal{L}_{\text{CE}} + \lambda_{OT} \mathcal{L}_{\text{OT}}
\end{equation}
\section{experiments}

\subsection{Experimental settings}
\subsubsection{Datasets}
We utilize the AudioCaps dataset~\cite{kim2019audiocaps} for training and evaluation, where each 10-second clip is annotated based on its audio component. Due to the limited accessibility of some YouTube links, we acquired 48,595 clips for the training and 944 clips for the test set. 
For audio pre-processing, we applied Short-Time Fourier Transform using a 25-ms Hanning window with a 10-ms hop size to each 10-second waveform sampled at 16 kHz, resulting in a $1024\times64$ spectrogram.
For visual input, 20 frames are uniformly selected from a 10-second clip at 2 FPS, then center-cropped to $224\times224$ pixels and normalized.

\begin{table}[t!]
\setlength{\tabcolsep}{10pt}
    \vspace{-1em}
    \caption{Ablation on audio-visual fusion method. $\mathcal{L}_\text{OT}$ is all applied.}
    \vspace{-0.5em}
    \centering
    \begin{adjustbox}{max width=\columnwidth}
    \begin{tabular}{lcccccc}
        \toprule
        Method & METEOR $\uparrow$ & CIDEr $\uparrow$ & SPICE $\uparrow$ & SPIDEr $\uparrow$ \\
        \midrule
        Q-Former  & 24.4 & 77.0 & 18.0 & 47.5  \\
        Joint encoder   & 24.9 & 76.7 & 18.4 &  47.5 \\
        Cross attention     & 24.8 & 80.1 & 17.9 & 49.0  \\
        OT-Att (ours)        & \textbf{26.2} & \textbf{84.9} & \textbf{18.5} & \textbf{51.7} \\
        \bottomrule
    \label{tab:abl_fusion}
  \end{tabular}
  \end{adjustbox}
  \vspace{-1.5em}
\end{table}
\begin{table}[t!]
\setlength{\tabcolsep}{10pt}
    \vspace{-1em}
    \caption{Ablation of the encoder and decoder training strategies.}
    \vspace{-0.5em}
    \centering
    \begin{adjustbox}{max width=\columnwidth}
    \begin{tabular}{cccccccc}
        \toprule
        Encoder & Decoder & METEOR $\uparrow$ & CIDEr $\uparrow$ & SPICE $\uparrow$ & SPIDEr $\uparrow$ \\
        \midrule
        Finetune & Freeze   & 25.3 & 77.7 & 17.3 & 47.5  \\
        LoRA & Freeze       & 24.4  & 73.5 & 17.3 & 45.4  \\
        Finetune & LoRA     & 25.5 & 78.9 & 18.1 & 48.5 \\
        LoRA &   LoRA       & \textbf{26.2} & \textbf{84.9} & \textbf{18.5} & \textbf{51.7} \\
        \bottomrule
    \label{tab:abl_freeze}
  \end{tabular}
  \end{adjustbox}
  \vspace{-2em}
\end{table}

\subsubsection{Implementation Details and Metrics}
During training, we utilize the AdamW optimizer with $\beta_1 = 0.9$, $\beta_2 = 0.999$, and a weight decay of 1e-6. For the first two epochs, out of a total of 100, the learning rate warms up to 5e-6, and then it gradually decreases following a cosine annealing strategy. 
We adopt a pre-trained Consistent Ensemble Distillation model~\cite{dinkel2024ced} as an audio encoder, and a pre-trained CLIP ViT-L/14 model~\cite{radford2021learning} as a visual encoder.
For text decoding, Llama 2~\cite{touvron2023llama} with 7B parameters is employed.
Both the audio encoder and text decoder are fine-tuned using low-rank adaptation (LoRA)~\cite{hu2022lora}, while the visual encoder is kept frozen.
When evaluating our methods, we use the metrics commonly employed for AAC, including BLEU~\cite{papineni2002bleu}, ROUGE-L~\cite{lin2004rouge}, METEOR~\cite{banerjee2005meteor}, CIDEr~\cite{vedantam2015cider}, SPICE~\cite{anderson2016spice}, and SPIDEr~\cite{liu2017improved}. Experiments are conducted using Intel Gaudi 2 AI Accelerator.

\subsection{Main Results}
The audio captioning performance of LAVCap on the AudioCaps dataset is shown in Table~\ref{tab:main_tab}. 
LAVCap not only outperforms previous works in closely matching the lexical content of ground truth captions but also shows enhanced semantic relevance and informativeness.
It is impressive that our framework achieves high performance although the ground-truth captions in AudioCaps are annotated based solely on audio.
Notably, our model still achieves comparable performance to the concurrent works~\cite{liu2024enhancing,haji2024taming} that utilize multiple and additional datasets for pre-training.
The results demonstrate that using an OT mapping between audio and visual modalities enables the model to train more effectively on semantically aligned features across these modalities, compared to previous methods such as cross-attention and concatenation. 
A further analysis of the results can be found in Section~\ref{sec/abl}.

\begin{table}[t!]
\setlength{\tabcolsep}{10pt}
    \caption{Examples of instruction prompts for the proposed method.}
    \vspace{-0.5em}
    \centering
    \begin{adjustbox}{max width=\columnwidth}
    \begin{tabular}{l}
        \toprule
        \small\textbf{\textit{Prompts I}} \\
        \textit{USER:} $<$Audio$>$\textbf{[Token$_{\textbf{a+v}}$]}$<$/Audio$>$ Please describe the audio.$\backslash n$ \\ \textit{ASSISTANT:} \\
        \midrule
        \small\textbf{\textit{Prompts II}} \\
        \textit{USER:} $<$Content$>$\textbf{[Token$_{\textbf{a+v}}$]}$<$/Content$>$ Please describe the content.$\backslash n$ \\ \textit{ASSISTANT:} \\
        \midrule
        \small\textbf{\textit{Prompts III}} \\
        \textit{USER:} $<$Audio$>$\textbf{[Tokens$_{\textbf{a}}$]}$<$/Audio$>$$<$Visual$>$\textbf{[Token$_{\textbf{v}}$]}$<$/Visual$>$ \\Please describe the audio and visual elements.$\backslash n$ \textit{ASSISTANT:} \\
        \bottomrule
    \label{tab:abl_prompt}
  \end{tabular}
  \end{adjustbox}
  \vspace{-2.2em}
\end{table}
\begin{table}[t!]
\setlength{\tabcolsep}{10pt}
    \caption{Ablation results based on examples of instruction prompts.}
    \vspace{-0.5em}
    \centering
    \begin{adjustbox}{max width=\columnwidth}
    \begin{tabular}{ccccccc}
        \toprule
        Prompt & METEOR $\uparrow$ & CIDEr $\uparrow$ & SPICE $\uparrow$ & SPIDEr $\uparrow$ \\
        \midrule
        \textit{Prompts I}  &  \textbf{26.2} & \textbf{84.9} & 18.5 & \textbf{51.7} \\
        \textit{Prompts II}        &  25.8 & 82.5 & 19.0 & 50.8  \\
        \textit{Prompts III}       &  25.9 & 81.8 & \textbf{19.1} & 50.5  \\
        \bottomrule
    \label{tab:abl_prompt_result}
  \end{tabular}
  \end{adjustbox}
  \vspace{-2.6em}
\end{table}

\subsection{Ablation Studies}
\label{sec/abl}

\subsubsection{Visual Modality and OT Loss}

We conduct an ablation study to evaluate the effectiveness of leveraging visual modality and OT loss.
As shown in Table.~\ref{tab:abl_split}, the performance improvement from simply adding visual modality is marginal without OT loss.
This demonstrates that bridging the modality gap through OT loss is crucial for enabling LLM to comprehend multi-modal contexts and process audio-visual features.

\subsubsection{Audio-Visual Fusion Methods}
Based on the utilization of visual modality and OT loss, we explore various audio-visual fusion methods. 
The results in Table.~\ref{tab:abl_fusion} show that the cross-modal fusion methods employed by the previous works are not effective in our setting.
We infer that this is due to the lack of training data to optimize the learnable parameters of these fusion methods and simultaneously align the fused audio-visual features to the LLM latent space.
On the other hand, the proposed OT attention module does not need any learnable parameters and leverages an optimal transport assignment map as an attention weight, thus providing effective audio-visual fusion in a data-efficient way.

\subsubsection{Encoder-Decoder Training Strategy}
After deciding on the optimal network architecture, we seek the best training strategy for our framework.
As shown in Table.~\ref{tab:abl_freeze}, training the encoder with LoRA instead of fine-tuning the entire parameters is much more efficient under data-limited circumstances.
In addition, freezing the weights of the LLM decoder severely degrades the performance, indicating that the LLM decoder should be adapted to the target dataset.

\subsubsection{Instruction Prompts}
To ensure that the LLM text decoder understands the input tokens properly, we explore various instruction prompts detailed in Table~\ref{tab:abl_prompt}. While \textit{prompts I} and \textit{II} handle audio and visual tokens together, \textit{prompt III} is specifically designed to understand them separately. Since AAC focuses primarily on audio-based captioning, \textit{prompt I} performs slightly better than the others, as shown in Table~\ref{tab:abl_prompt_result}.

\begin{figure}[t!]
    \centering
    \label{fig:qual_1}
    \includegraphics[width=0.95\columnwidth]{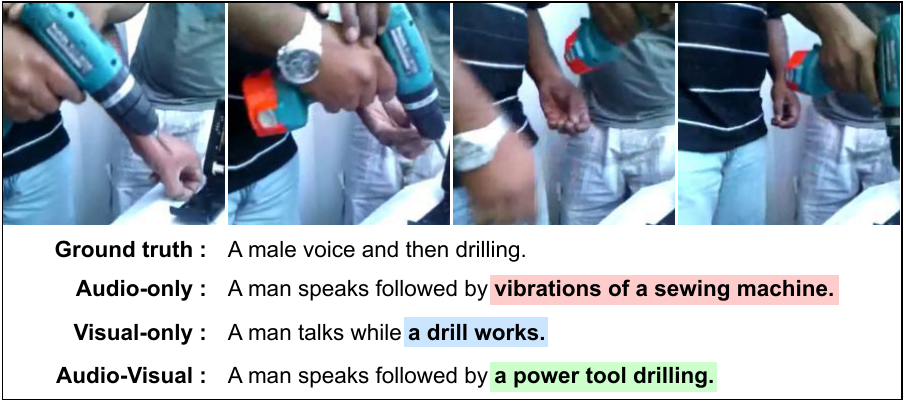}   
    \vspace{0.2em}
    \centering
    \label{fig:qual_2}
    \includegraphics[width=0.95\columnwidth]{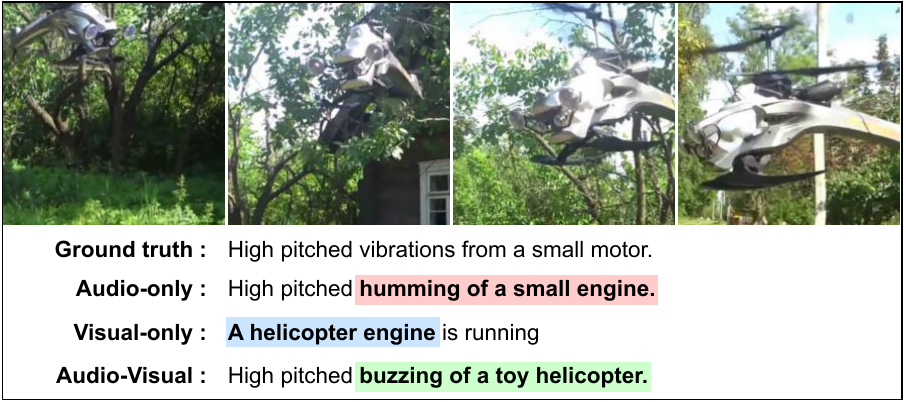}
    \vspace{-0.7em}
    \caption{Qualitative results of captions generated from models trained solely on audio, only on visual, and on both audio and visual.}
    \vspace{-0.8em}
    \label{fig:qual}
\end{figure}

\begin{table}[t]
\setlength{\tabcolsep}{10pt}
    \caption{MOS with 95$\%$ confidence intervals.}
    \vspace{-0.5em}
    \centering
    \setlength{\tabcolsep}{0.08\columnwidth}
    \begin{tabular}{lc}
        \toprule
        Input modality & MOS $\uparrow$ (1$\sim$5) \\
        \midrule
        ground truth            & 3.81 $\pm$ 0.12 \\
        audio-only              & 2.89 $\pm$ 0.12 \\
        visual-only             & 2.71 $\pm$ 0.12 \\
        audio-visual     & \textbf{4.08} $\pm$ \textbf{0.10} \\
        \bottomrule
    \label{tab:mos}
  \end{tabular}
  \vspace{-2.8em}
\end{table}

\subsection{Qualitative Results}
In Fig.~\ref{fig:qual} we visualize the captions generated by models trained with audio-only, visual-only, and combined audio-visual data. 
While using a single modality may capture incorrect elements, utilizing both audio and visual inputs provides a more detailed view of the video. 
Furthermore, we conduct a user study with 20 participants, using MOS to rate each of the 35 generated captions from 1 to 5 based on how well they describe the video.
The results in Table~\ref{tab:mos} show that the captions generated by LAVCap concretely represent the audio content.
Remarkably, MOS of LAVCap is even higher than the ground-truth captions.
This highlights the significant benefit of incorporating the visual modality to better distinguish the various sounds and comprehend the scene context.
\section{conclusion}
In this work, we propose LAVCap, an LLM-based audio-visual captioning framework designed to incorporate visual information into the audio modality using an optimal transport-based strategy. 
Specifically, \textit{OT loss} and \textit{OT-Att} are introduced to align the modality gap and promote effective fusion of audio and visual features.
The proposed model outperforms previous captioning methods on the AudioCaps dataset, without the need for pre-training on large datasets or post-processing, highlighting its promise in designing new audio-visual fusion methods. 

\section{Acknowledgment}
This research was supported in part by the NAVER-Intel Co-Lab. The work was conducted by Korea Advanced Institute of Science and Technology and reviewed by both NAVER and Intel.

\bibliographystyle{ieeetr}
\bibliography{biblio}

\begin{thebibliography}{10}

\bibitem{huang2023personalized}
Q.~Huang, Y.~Zhang, T.~Ko, X.~Liu, B.~Wu, W.~Wang, and H.~Tang, ``Personalized dialogue generation with persona-adaptive attention,'' in {\em Proceedings of the AAAI Conference on Artificial Intelligence}, pp.~12916--12923, 2023.

\bibitem{mei2021act}
X.~Mei, X.~Liu, Q.~Huang, M.~D. Plumbley, and W.~Wang, ``Audio captioning transformer,'' in {\em Proceedings of the 6th Detection and Classification of Acoustic Scenes and Events 2021 Workshop}, pp.~211--215, 2021.

\bibitem{gontier2021barttags}
F.~Gontier, R.~Serizel, and C.~Cerisara, ``Automated audio captioning by fine-tuning bart with audioset tags,'' in {\em DCASE 2021 - 6th Workshop on Detection and Classification of Acoustic Scenes and Events}, 2021.

\bibitem{kim2023prefix}
M.~Kim, K.~Sung-Bin, and T.-H. Oh, ``Prefix tuning for automated audio captioning,'' in {\em International Conference on Acoustics, Speech and Signal Processing}, pp.~1--5, 2023.

\bibitem{kim2024enclap}
J.~Kim, J.~Jung, J.~Lee, and S.~H. Woo, ``{EnCLAP}: Combining neural audio codec and audio-text joint embedding for automated audio captioning,'' in {\em International Conference on Acoustics, Speech and Signal Processing}, pp.~6735--6739, 2024.

\bibitem{mei2024wavcaps}
X.~Mei, C.~Meng, H.~Liu, Q.~Kong, T.~Ko, C.~Zhao, M.~D. Plumbley, Y.~Zou, and W.~Wang, ``{W}av{C}aps: A chatgpt-assisted weakly-labelled audio captioning dataset for audio-language multimodal research,'' {\em IEEE/ACM Transactions on Audio, Speech, and Language Processing}, 2024.

\bibitem{labb2024conette}
E.~Labb, T.~Pellegrini, J.~Pinquier, {\em et~al.}, ``{CoNeTTE}: An efficient audio captioning system leveraging multiple datasets with task embedding,'' {\em IEEE/ACM Transactions on Audio, Speech, and Language Processing}, 2024.

\bibitem{liu2024enhancing}
J.~Liu, G.~Li, J.~Zhang, H.~Dinkel, Y.~Wang, Z.~Yan, Y.~Wang, and B.~Wang, ``Enhancing automated audio captioning via large language models with optimized audio encoding,'' in {\em Interspeech}, pp.~1135--1139, 2024.

\bibitem{haji2024taming}
M.~Haji-Ali, W.~Menapace, A.~Siarohin, G.~Balakrishnan, S.~Tulyakov, and V.~Ordonez, ``Taming data and transformers for audio generation,'' {\em arXiv preprint arXiv:2406.19388}, 2024.

\bibitem{kong2020panns}
Q.~Kong, Y.~Cao, T.~Iqbal, Y.~Wang, W.~Wang, and M.~D. Plumbley, ``{PANNs}: Large-scale pretrained audio neural networks for audio pattern recognition,'' {\em IEEE/ACM Transactions on Audio, Speech, and Language Processing}, vol.~28, pp.~2880--2894, 2020.

\bibitem{chen2022hts}
K.~Chen, X.~Du, B.~Zhu, Z.~Ma, T.~Berg-Kirkpatrick, and S.~Dubnov, ``{HTS-AT}: A hierarchical token-semantic audio transformer for sound classification and detection,'' in {\em International Conference on Acoustics, Speech and Signal Processing}, pp.~646--650, 2022.

\bibitem{chen2023beats}
S.~Chen, Y.~Wu, C.~Wang, S.~Liu, D.~Tompkins, Z.~Chen, W.~Che, X.~Yu, and F.~Wei, ``{BEATs}: audio pre-training with acoustic tokenizers,'' in {\em Proceedings of the 40th International Conference on Machine Learning}, 2023.

\bibitem{elizalde2023clap}
B.~Elizalde, S.~Deshmukh, M.~Al~Ismail, and H.~Wang, ``{CLAP}: Learning audio concepts from natural language supervision,'' in {\em International Conference on Acoustics, Speech and Signal Processing}, pp.~1--5, 2023.

\bibitem{défossez2023high}
A.~D{\'e}fossez, J.~Copet, G.~Synnaeve, and Y.~Adi, ``High fidelity neural audio compression,'' {\em Transactions on Machine Learning Research}, 2023.

\bibitem{liu2023visually}
X.~Liu, Q.~Huang, X.~Mei, H.~Liu, Q.~Kong, J.~Sun, S.~Li, T.~Ko, Y.~Zhang, L.~H. Tang, {\em et~al.}, ``Visually-aware audio captioning with adaptive audio-visual attention,'' in {\em Interspeech}, pp.~2838--842, 2023.

\bibitem{kim2024avcap}
J.~Kim, J.~Shin, and J.~Kim, ``{AVCap}: Leveraging audio-visual features as text tokens for captioning,'' in {\em Interspeech}, 2024.

\bibitem{lin2024norton}
Y.~Lin, J.~Zhang, Z.~Huang, J.~Liu, Z.~Wen, and X.~Peng, ``Multi-granularity correspondence learning from long-term noisy videos,'' in {\em The Twelfth International Conference on Learning Representations}, 2024.

\bibitem{cuturi2013sinkhorn}
M.~Cuturi, ``Sinkhorn distances: Lightspeed computation of optimal transport,'' in {\em Proceedings of the 26th International Conference on Neural Information Processing Systems}, p.~2292–2300, 2013.

\bibitem{kim2019audiocaps}
C.~D. Kim, B.~Kim, H.~Lee, and G.~Kim, ``Audiocaps: Generating captions for audios in the wild,'' in {\em Proceedings of the 2019 Conference of the North American Chapter of the Association for Computational Linguistics: Human Language Technologies, Volume 1 (Long and Short Papers)}, pp.~119--132, 2019.

\bibitem{dinkel2024ced}
H.~Dinkel, Y.~Wang, Z.~Yan, J.~Zhang, and Y.~Wang, ``Ced: Consistent ensemble distillation for audio tagging,'' in {\em International Conference on Acoustics, Speech and Signal Processing}, pp.~291--295, 2024.

\bibitem{radford2021learning}
A.~Radford, J.~W. Kim, C.~Hallacy, A.~Ramesh, G.~Goh, S.~Agarwal, G.~Sastry, A.~Askell, P.~Mishkin, J.~Clark, {\em et~al.}, ``Learning transferable visual models from natural language supervision,'' in {\em Proceedings of the 38th International Conference on Machine Learning}, pp.~8748--8763, 2021.

\bibitem{touvron2023llama}
H.~Touvron, L.~Martin, K.~Stone, P.~Albert, A.~Almahairi, Y.~Babaei, N.~Bashlykov, S.~Batra, P.~Bhargava, S.~Bhosale, {\em et~al.}, ``Llama 2: Open foundation and fine-tuned chat models,'' {\em arXiv preprint arXiv:2307.09288}, 2023.

\bibitem{hu2022lora}
E.~J. Hu, Y.~Shen, P.~Wallis, Z.~Allen-Zhu, Y.~Li, S.~Wang, L.~Wang, and W.~Chen, ``Lo{RA}: Low-rank adaptation of large language models,'' in {\em The Tenth International Conference on Learning Representations}, 2022.

\bibitem{papineni2002bleu}
K.~Papineni, S.~Roukos, T.~Ward, and W.-J. Zhu, ``{BLEU}: a method for automatic evaluation of machine translation,'' in {\em Proceedings of the 40th Annual Meeting of the Association for Computational Linguistics}, p.~311–318, 2002.

\bibitem{lin2004rouge}
C.-Y. Lin, ``{ROUGE}: A package for automatic evaluation of summaries,'' in {\em Text Summarization Branches Out}, pp.~74--81, 2004.

\bibitem{banerjee2005meteor}
S.~Banerjee and A.~Lavie, ``{METEOR}: An automatic metric for {MT} evaluation with improved correlation with human judgments,'' in {\em Proceedings of the ACL Workshop on Intrinsic and Extrinsic Evaluation Measures for Machine Translation and/or Summarization}, pp.~65--72, 2005.

\bibitem{vedantam2015cider}
R.~Vedantam, C.~Lawrence~Zitnick, and D.~Parikh, ``Cider: Consensus-based image description evaluation,'' in {\em Proceedings of the IEEE/CVF Conference on Computer Vision and Pattern Recognition}, pp.~4566--4575, 2015.

\bibitem{anderson2016spice}
P.~Anderson, B.~Fernando, M.~Johnson, and S.~Gould, ``Spice: Semantic propositional image caption evaluation,'' in {\em Proceedings of the European Conference on Computer Vision}, pp.~382--398, 2016.

\bibitem{liu2017improved}
S.~Liu, Z.~Zhu, N.~Ye, S.~Guadarrama, and K.~Murphy, ``Improved image captioning via policy gradient optimization of spider,'' in {\em Proceedings of the IEEE/CVF International Conference on Computer Vision}, pp.~873--881, 2017.

\end{thebibliography}

\end{document}